\documentclass[11pt]{article}
\usepackage[letterpaper,margin=1in]{geometry}

\usepackage{amsmath}
\usepackage{amssymb}
\usepackage{amsthm}
\usepackage{graphicx}
\usepackage{booktabs}
\usepackage{enumitem}
\usepackage{xcolor}
\usepackage{tikz}
\usepackage{eso-pic}
\usepackage{url}
\usetikzlibrary{positioning,shapes.geometric,arrows.meta}

\providecommand{\citenamefont}[1]{#1}

\newtheorem{theorem}{Theorem}[section]

\newtheorem{definition}[theorem]{Definition}

\newtheorem{principle}[theorem]{Principle}

\newcommand{\fito}{\textsc{fito}}
\newcommand{\oae}{\textsc{oae}}
\newcommand{\rdma}{\textsc{rdma}}
\newcommand{\pif}{\textsc{pif}}
\newcommand{\kbp}{\textsc{kbp}}
\newcommand{\rcp}{\textsc{rcp}}
\newcommand{\ico}{\textsc{ico}}
\newcommand{\ac}{\textsc{ac}}
\newcommand{\lww}{\textsc{lww}}

\usepackage{natbib}

\title{\textbf{The Semantic Arrow of Time, Part~V:}\\[0.3em]{\large The Leibniz Bridge --- Toward a Unified Theory of Semantic Time}}
\author{Paul Borrill \\ D\AE D\AE LUS}
\date{02026-FEB-27}

\usepackage[hidelinks]{hyperref}

\begin{document}
\maketitle

\begin{center}
\large\itshape Toward a Unified Theory of Semantic Time
\end{center}
\vspace{1em}

\begin{abstract}
\noindent
This is the final paper in the five-part series \emph{The Semantic
Arrow of Time}.  Part~I~\citep{borrill2026-partI} identified the
\fito{} category mistake---treating forward temporal flow as
sufficient for establishing meaning.
Part~II~\citep{borrill2026-partII} presented the constructive
alternative: the \oae{} link state machine with its mandatory
reflecting phase.  Part~III~\citep{borrill2026-partIII} showed the
\fito{} fallacy operating at industrial scale in \rdma{} completion
semantics.  Part~IV~\citep{borrill2026-partIV} traced the same
pattern through file synchronization, email, human memory, and
language model hallucination.

This paper closes the series by constructing the \emph{Leibniz
Bridge}: a unified framework that connects the philosophical
foundations (Leibniz's Identity of Indiscernibles, as formalized by
Spekkens), the protocol engineering (\oae{}'s bilateral transaction
structure), and the physical substrate (indefinite causal order in
quantum mechanics).  The bridge rests on a single principle:
\emph{mutual information conservation}---the requirement that every
causal exchange preserve the total information accessible to both
endpoints, with the direction of time emerging not from axiom but
from entropy production when a reversible exchange commits.

We show that this principle dissolves the apparent impossibility of
the FLP, Two Generals, and CAP theorems by revealing them as
theorems about \fito{} systems, not about physics.  We present the
triangle network as the minimal topology for semantic consistency
without centralized coordination.  We conclude with open questions
and a reflection on what distributed computing looks like when the
\fito{} assumption is dropped.
\end{abstract}

\section[Introduction]{Introduction: Five Papers, One Mistake}
\label{sec:intro}

This series has told a single story through five different lenses.
The story is that computing inherited an assumption about time---that
causation flows forward, that temporal succession implies semantic
progress, that the return path is overhead rather than constitutive---and
that this assumption, which we have called Forward-In-Time-Only
(\fito{}), is a category mistake in the precise sense that Gilbert
Ryle~\citep{ryle1949} defined the term: treating a concept that
belongs to one logical type as if it belongs to another.

Part~I traced the mistake from Eddington's 1927 coinage of ``arrow of
time'' through Shannon's unidirectional channel model and Lamport's
happened-before relation, showing that physics has begun to question
the default assumption while computing has not.  Part~II presented
the \oae{} link state machine as a protocol that \emph{constructs}
causal order through bilateral exchange rather than \emph{assuming}
it through timestamps.  Part~III documented the consequences when
\fito{} is implemented at hardware scale: \rdma{}'s completion signal
reports success at temporal stage~$T_4$ while semantic agreement
remains unestablished at $T_6$.  Part~IV showed that the same
pattern---forward commitment without reflection---produces silent data
destruction in cloud sync, phantom messages in email, false memories
in human cognition, and hallucinations in language models.

The question this final paper must answer is: \emph{what replaces
\fito{}?}  Not merely at the protocol level (Part~II answered that)
or at the diagnostic level (Parts~III and~IV answered that), but at
the level of principle.  What is the foundational commitment that,
once adopted, makes the \fito{} category mistake impossible to make?

The answer is the Leibniz Bridge.

\section[The Leibniz Principle]{The Leibniz Principle in Three Communities}
\label{sec:leibniz}

Gottfried Wilhelm Leibniz formulated the Identity of Indiscernibles
in 1686: if two entities share all the same properties, they are not
two entities but one.  Rob Spekkens~\citep{spekkens2019} elevated
this from a metaphysical axiom to a \emph{methodological principle
for theory construction}:

\begin{quote}
\emph{If a theory predicts two scenarios that are ontologically
distinct but empirically indiscernible, the theory should be rejected
in favor of one where the scenarios are ontologically identical.}
\end{quote}

Spekkens showed that Einstein applied this principle repeatedly:
eliminating absolute simultaneity (special relativity), eliminating
the distinction between inertial and gravitational mass (general
relativity).  Each breakthrough came not from new experimental
evidence but from recognizing that an ontological distinction without
empirical consequence was a sign of \emph{surplus structure} in the
theory.%

The claim of this series is that the same surplus structure pervades
three communities, and that the Leibniz principle dissolves it in
each case.

\subsection{Computer Science: The Missing Fourth Cell}

Lamport's happened-before relation~\citep{lamport1978} defines a
partial order on events.  For any pair of events $a$ and $b$, exactly
one of three relations holds: $a \to b$ (``$a$ happened before
$b$''), $b \to a$, or $a \| b$ (concurrent: no causal connection).
Vector clocks, logical clocks, and their descendants all inherit this
three-valued framework.

But a 2-bit encoding of causal relations has four cells, not three:

\begin{center}
\small
\begin{tabular}{ccl}
\toprule
\textbf{Encoding} & \textbf{Relation} & \textbf{Framework} \\
\midrule
01 & $a \to b$ & Definite causal order (Lamport) \\
10 & $b \to a$ & Definite causal order (Lamport) \\
00 & $a \| b$ (concurrent) & Partial order (vector clocks) \\
11 & $a \leftrightarrow b$ (indefinite) & \textbf{No existing CS framework} \\
\bottomrule
\end{tabular}
\end{center}

The fourth cell---encoding~11, representing events connected by a
bilateral exchange whose causal direction is genuinely
\emph{indefinite}---has no representation in any existing computer
science framework.  This is the gap.  Indefinite Logical Timestamps
and their generalization, Tensor Clocks, fill this gap by providing
a four-valued causal algebra that includes the indefinite
relation~\citep{borrill2026-partII}.%

\subsection{Networking: The Return Path Is Constitutive}

Every real Ethernet link is physically bidirectional.  A transmitter
sends data; a receiver acknowledges.  But protocol design treats the
return path as overhead---the ACK is a reliability mechanism, not a
constitutive feature of the channel.  Shannon's original
model~\citep{shannon1948} is unidirectional: $X \to Y$, with
$I(X;Y) = H(X) - H(X|Y)$.

The Leibniz principle reveals the surplus structure: if both
directions of a link carry independent information and are
empirically indistinguishable before the protocol designer labels
one ``forward,'' then the distinction between forward and backward
is not a property of the link but of the model.  The
\emph{direction of time} on a bilateral link is imposed by the
protocol designer, not by the physics of the medium.%

Perfect Information Feedback (\pif{}) models the link as two
conjugate Shannon channels forming a closed informational system.
The static symmetry is:
\begin{equation}
  I_+(X;Y) = I_-(Y;X)
  \label{eq:pif-symmetry}
\end{equation}
This is a property of the joint distribution $P(X,Y)$, not a
dynamical flow.  As Lee observes~\citep{lee-email2026}, in Shannon's
formulation mutual information is not a function of time---it gives the
information conveyed \emph{per use of the channel}, a theoretical
limit requiring perfect codes and unbounded delay.  It says nothing
about timing.%

The operational consequence for \oae{} is that a bilateral link's
capacity is $C_{\pif{}} = 2 \cdot C_{\text{one-way}}$---not because
it sends data in both directions, but because the echo \emph{confirms}
the forward assertion and closes the causal
loop~\citep{munamala2025}.  The two conjugate channels provide
complementary views of the same mutual information: the ``orientation''
of arriving versus departing packets is a labelling convention, not a
physical asymmetry~\citep{stone2018}.

\subsection{Physics: Indefinite Causal Order}

Oreshkov, Costa, and Brukner~\citep{oreshkov2012} proved that
quantum mechanics is consistent with processes that have no definite
causal order.  Rubino et al.~\citep{rubino2017} verified this
experimentally using the quantum switch.  Zhao et al.~\citep{zhao2025}
demonstrated that \ico{} provides a genuine communication
advantage: the quantum communication power of indefinite causal
order is strictly greater than any fixed-order protocol.

The Leibniz principle completes the picture.  Bell's resolution of
EPR---accepting non-locality while maintaining \fito{}---is, as
Spekkens and Lee independently observe, a \emph{choice}, not a
theorem.  The phenomena that physics considers exotic (\ico{},
process matrices, causal superposition) have been hiding in plain
sight in the most basic networking primitive: the acknowledged
packet on a bidirectional link.

\section[Mutual Information Conservation]{The Bridge: Mutual Information Conservation}
\label{sec:conservation}

The Leibniz Bridge is the principle that connects the philosophical
foundation (Leibniz--Spekkens), the protocol engineering (\oae{}),
and the physical substrate (\ico{}).  We state it as a formal
principle:

\begin{principle}[Mutual Information Conservation]
\label{prin:mic}
For any bilateral causal exchange between endpoints $A$ and $B$:
\begin{enumerate}[leftmargin=1cm, nosep]
  \item The total mutual information $I(A;B)$ accessible to the
    joint system is conserved throughout the exchange.
  \item The direction of information flow (the arrow of time on
    the link) is not a property of the physical exchange but of
    the entropy production when the exchange commits.
  \item Any protocol that signals completion before mutual
    information conservation is verified violates the Leibniz
    principle and will produce semantic corruption.
\end{enumerate}
\end{principle}

This principle has immediate consequences for each of the domains
examined in this series:

\paragraph{For protocol design (Part~II):}
The \oae{} link state machine's mandatory \textsc{reflecting} phase
is not an engineering choice but a \emph{conservation requirement}.
The reflecting phase is where the receiver communicates back to the
sender what it actually received---closing the informational loop
and verifying that $I(A;B)$ has been preserved.  Without this phase,
the sender has no evidence that the information it transmitted
arrived with its meaning intact.

\begin{figure*}[h]
\centering
\includegraphics[width=\textwidth]{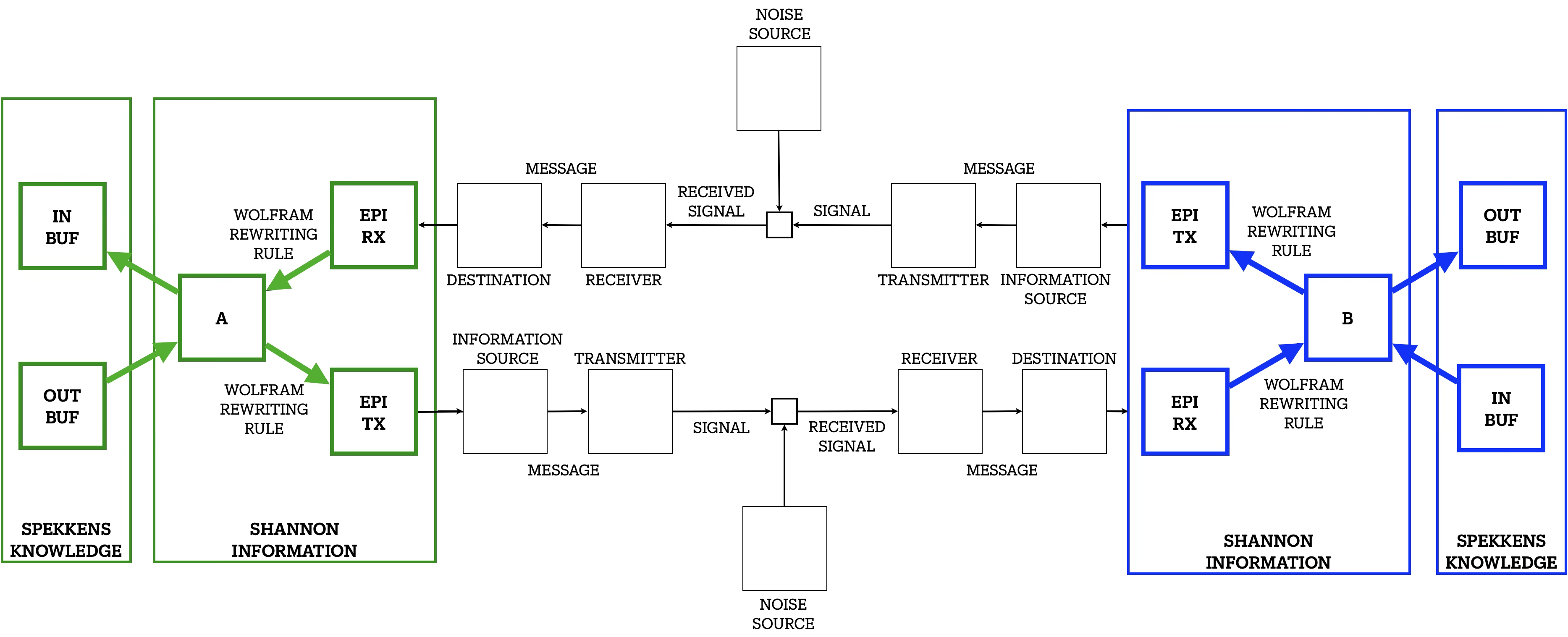}
\caption{The Leibniz Bridge architecture.  Two conjugate Shannon
channels (centre) connect endpoints $A$ (green) and $B$ (blue),
each equipped with Spekkens Knowledge registers (EPI~TX, EPI~RX)
and Wolfram rewriting rules that enforce the Knowledge Balance
Principle.  The IN/OUT buffers at each endpoint hold the ontic
state; the EPI registers hold the epistemic state---exactly half,
per the \kbp{}.  Noise sources act on each channel independently.
The bilateral structure makes the return path constitutive, not
auxiliary: information conservation is verified through the closed
loop, not assumed from forward delivery alone.}
\label{fig:general-comm}
\end{figure*}

\paragraph{For \rdma{} (Part~III):}
The completion fallacy is a violation of Principle~\ref{prin:mic}:
the CQE at $T_4$ signals completion before the receiver has verified
information conservation.  The gap between $T_4$ and $T_6$ is the
space in which mutual information is potentially lost---and the
completion signal's claim of success prevents the sender from
detecting the loss.

\paragraph{For file sync and email (Part~IV):}
\lww{} conflict resolution violates Principle~\ref{prin:mic} by
discarding one side of a concurrent modification without verifying
that the discarded information is recoverable.  Timestamp-based
email ordering violates it by imposing a causal direction (``this
operation happened before that one'') without verifying that the
imposed direction preserves the user's intended meaning.

\paragraph{For memory and language models (Part~IV):}
Schema-driven encoding (Bartlett) and autoregressive generation
(transformers) both violate Principle~\ref{prin:mic} by committing
state changes forward without verifying that the committed state
preserves the information content of the original stimulus or
intended meaning.  Sleep consolidation and RLHF are partial
mitigation strategies that reduce but do not eliminate the violation.

\section[The Reversible Causal Principle]{The Reversible Causal Principle}
\label{sec:rcp}

The Alternating Causality (\ac{}) framework, developed in the
\emph{Entropy} pre-print~\citep{munamala2025}, consolidates the
mutual information conservation principle into a single symmetry law:
\begin{equation}
  P_{BA}(t) = P_{AB}(-t)^{\dagger}
  \label{eq:rcp}
\end{equation}
Every causal interaction $A \leftrightarrow B$ admits a time-reversed
conjugate.  Any closed causal loop satisfies $\oint_\Gamma d\varphi
= 0$ (zero net causal flux).  This extends Noether's theorem:

\begin{description}[leftmargin=1.5cm]
  \item[Noether (1918):] Time-symmetry $\implies$ energy conservation.
  \item[\rcp{}:] Causal-symmetry $\implies$ information conservation.
\end{description}

When composition is perfect (no noise, no loss), the reverse
operator $R$ satisfies $R \circ F = \mathbb{I}$, and information is
shuttled without destruction.  When composition is imperfect, the
deficit $\Lambda$ in $X^\dagger X = \mathbb{I} - \Lambda$ measures
\emph{informational entropy generation}---causal decoherence.  The
arrow of time \emph{emerges} from this entropy production, not from
axiom.%

The \rcp{} imposes conservation laws that mirror electrical circuit
theory---Kirchhoff laws for information:

\begin{description}[leftmargin=1.5cm]
  \item[Node law:] $\displaystyle\sum_{e \in \mathrm{in}(v)} J_e[n]
    = \sum_{e \in \mathrm{out}(v)} J_e[n]$\quad (information flux
    conserved at every node, every slot).
  \item[Loop law:] $\displaystyle\sum_{e \in \Gamma} \Delta\Phi_e[n]
    = 0$\quad (no net bias around any closed causal cycle).
\end{description}

These are not metaphors.  They are exact conservation laws for the
information potential $\Phi_e$ (surprisal, $-\log p$, or
log-likelihood ratio).  A \pif{} link behaves like a lossless AC
network at steady state; noise introduces resistive drops that
convert balanced standing waves into directed, entropy-producing
flow~\citep{munamala2025}.

\section[Knowledge Balance]{The Knowledge Balance Principle as Protocol Design}
\label{sec:kbp}

Spekkens' Knowledge Balance Principle (\kbp{})~\citep{spekkens2007}
states: \emph{For maximal knowledge of any system, what you know must
equal what you don't know.}  For a system with $2N$ bits of ontic
state, an epistemic agent can know at most $N$ bits.

This principle, which Spekkens developed for quantum foundations,
has a direct engineering realization in the \oae{} link
protocol~\citep{borrill2022}:

\begin{description}[leftmargin=1.5cm]
  \item[\textbf{ONT registers}:] Represent the \emph{imagined} ontic
    state of the link: 4~bits (2~bits per direction $\times$
    2~directions).  These registers are \textbf{not real}.  They are a
    theoretical construct---by definition, no single endpoint can ever
    access the full ontic state.  The ONT register is the state the
    link ``would have'' if an omniscient observer could see both sides
    simultaneously; no such observer exists in any physical
    implementation.%
  \item[\textbf{EPI registers}:] Hold the \emph{accessible} epistemic
    state: 2~bits (exactly half the ontic state, per \kbp{}).  This
    is all any endpoint can ever know about the link.
\end{description}

Each endpoint knows exactly half the link's ontic state---it knows
its own contribution but not the other side's.  This is not a
design limitation; it is a \emph{conservation law} applied to
knowledge.  The inaccessibility of the ONT register is not a
shortcoming to be engineered around---it is a \emph{constitutive
feature} of any bilateral exchange, as fundamental as the
uncertainty principle in quantum mechanics.  Six consistent Spekkens states map to valid protocol
states.  Thirty-six product states describe separable link
configurations.  Twenty-four ``entangled'' states describe non-local
correlations between endpoints.  The Wolfram rewriting rules on EPI
registers enforce state transitions that respect the knowledge
balance at every step~\citep{borrill2022}.%

The connection to Parts~III and~IV is immediate.  \rdma{}'s
completion signal gives the sender the illusion of knowing the full
state of the transaction---``the write succeeded''---when in fact it
knows only its own half (the data was placed).  The receiver's
half (the data was semantically integrated) is unknown and, under
\fito{} semantics, unknowable until the application performs its
own verification.  The \kbp{} formalizes this asymmetry and makes
it a design constraint: no protocol should signal completion unless
both halves of the knowledge balance are accounted for.

\section[Impossibility Theorems]{Dissolving the Impossibility Theorems}
\label{sec:impossibility}

Three impossibility results have shaped distributed systems theory
for decades: the FLP impossibility of consensus~\citep{fischer1985},
the Two Generals problem~\citep{gray1978}, and the CAP
theorem~\citep{brewer2000,gilbert2002}.  Each is treated as a
fundamental limitation of distributed computing.  The Leibniz Bridge
reveals them as theorems about \fito{} systems, not about physics.

\subsection{FLP: Consensus Under \fito{}}

Fischer, Lynch, and Paterson~\citep{fischer1985} proved that no
deterministic protocol can guarantee consensus in an asynchronous
system if even one process can fail.  The proof assumes that
processes communicate through a message system with \emph{one-way}
delivery: a process sends a message, and the message eventually
arrives, but the sender receives no confirmation of delivery timing.

This is \fito{} encoded as a system model.  The message system is
a Shannon channel: $X \to Y$, forward-only, with the return path
absent from the model.  Under these assumptions, FLP is correct:
you cannot distinguish a slow process from a dead one, and therefore
cannot decide when to commit.

Under the Leibniz Bridge, the question changes.  A bilateral
exchange with a \textsc{reflecting} phase does not need to
distinguish slow from dead---it needs only to verify that the
exchange completed within the transaction's entropy budget.  If the
reflection arrives, the exchange is committed.  If it does not
arrive within the link's temporal horizon, the exchange is aborted
and the state is rolled back to the last committed
checkpoint.%

\subsection{Two Generals: The Prototype \fito{} Impossibility}

The Two Generals problem~\citep{gray1978} is the simplest
impossibility result: two generals must agree on an attack time, but
every messenger may be captured.  No finite number of messages can
guarantee agreement, because neither general can know that their
last message was received.

This is \fito{} in its purest form: one-way messages, no reflecting
phase, no bilateral confirmation.  The problem assumes that
communication is fundamentally unidirectional---a messenger travels
from $A$ to $B$ or from $B$ to $A$, but never simultaneously in
both directions.

The \oae{} framework resolves this by changing the communication
primitive.  Instead of sending messengers, the generals share a
bilateral link that implements the TIKTYKTIK protocol: a four-message
reversible exchange that establishes common knowledge through
alternating causality, without requiring any individual message to
be guaranteed.  The protocol does not defeat the Two Generals
impossibility---it dissolves it by replacing the system model in
which the impossibility holds.

\subsection{CAP: Consistency, Availability, and \fito{}}

Brewer's CAP theorem~\citep{brewer2000}, formalized by Gilbert and
Lynch~\citep{gilbert2002}, states that a distributed system cannot
simultaneously provide Consistency, Availability, and Partition
tolerance.  The standard interpretation: during a network partition,
you must choose between consistency (refusing to serve stale data)
and availability (serving potentially inconsistent data).

The Leibniz Bridge reframes CAP.  In a \fito{} system, a partition
means that forward messages cannot reach their destination, and
therefore the system cannot maintain consistency.  In a bilateral
system with the \rcp{}, a partition means that the reflecting phase
cannot complete, which means the transaction cannot commit.  The
system does not serve inconsistent data---it holds the transaction
in the \textsc{tentative} state until the partition heals or an
alternative path is found.

The triangle network (Section~\ref{sec:triangle}) provides the
topological mechanism: when a direct link partitions, the third
vertex of every triangle provides an alternative reflecting path.
This does not eliminate partitions---physics guarantees they will
occur---but it transforms partition handling from a binary choice
(consistency or availability) into a topological property
(alternative reflecting paths).

\section[The Triangle Network]{The Triangle Network: Minimal Topology for Semantic Consistency}
\label{sec:triangle}

The \oae{} architecture proposes the triangle as the minimal topology
for semantic consistency without centralized coordination.  The
argument proceeds in three steps.

\subsection{Why Triangles}

\begin{figure}[h]
\centering
\includegraphics[width=0.19\columnwidth]{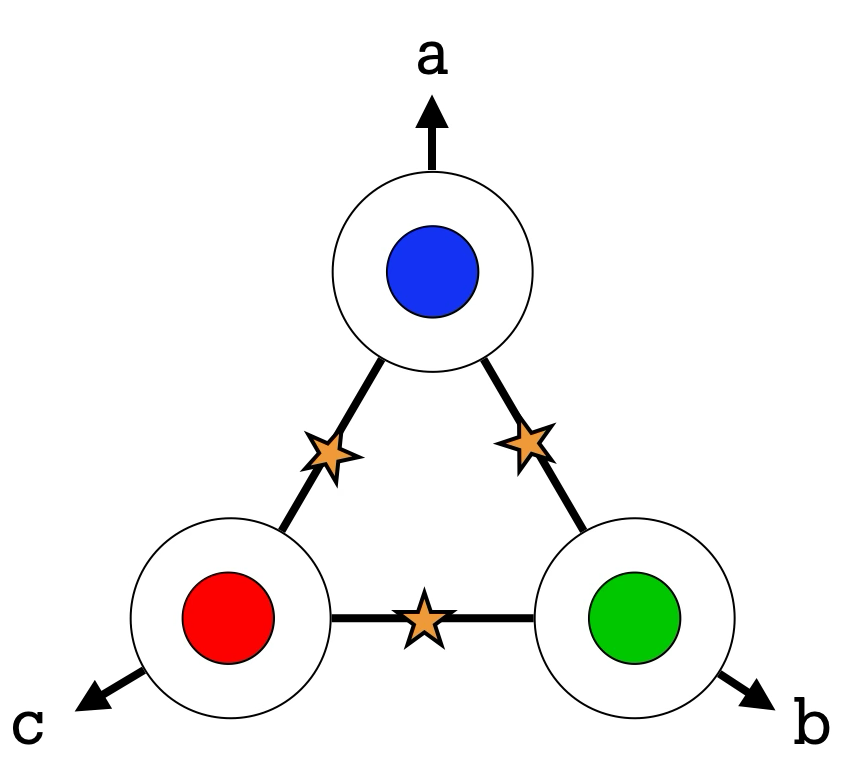}\hfill
\includegraphics[width=0.19\columnwidth]{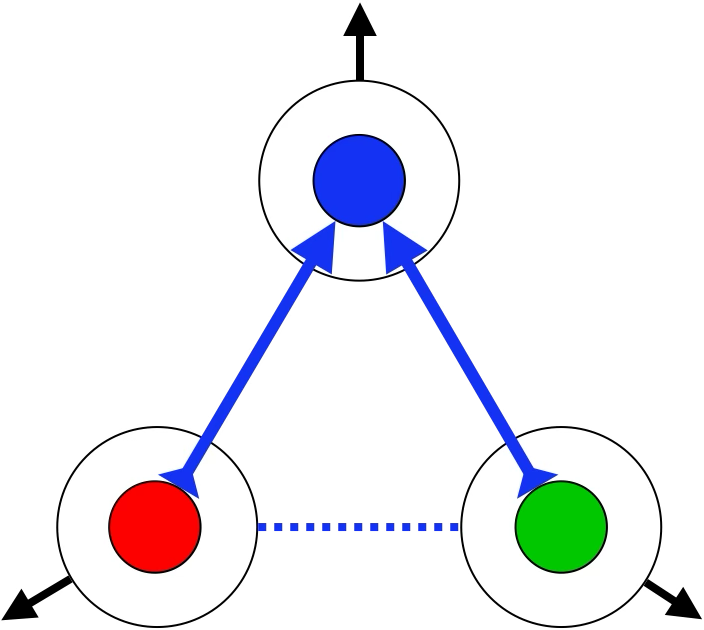}\hfill
\includegraphics[width=0.19\columnwidth]{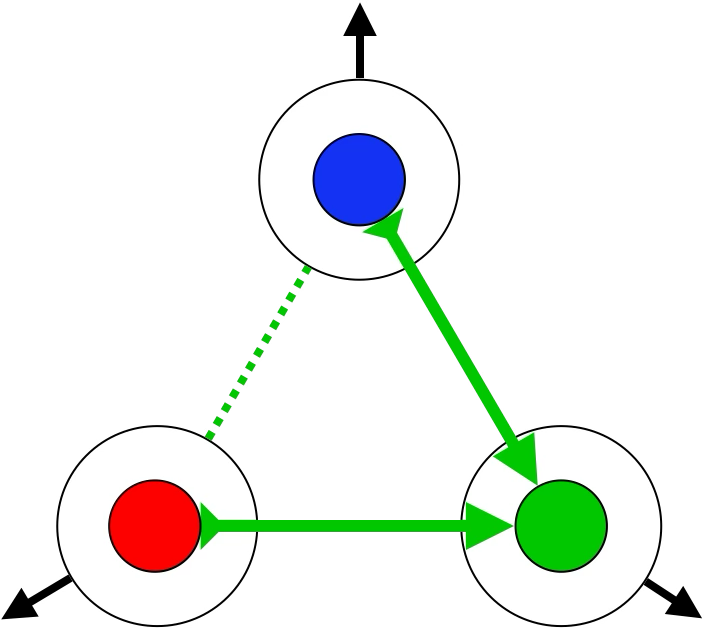}\hfill
\includegraphics[width=0.19\columnwidth]{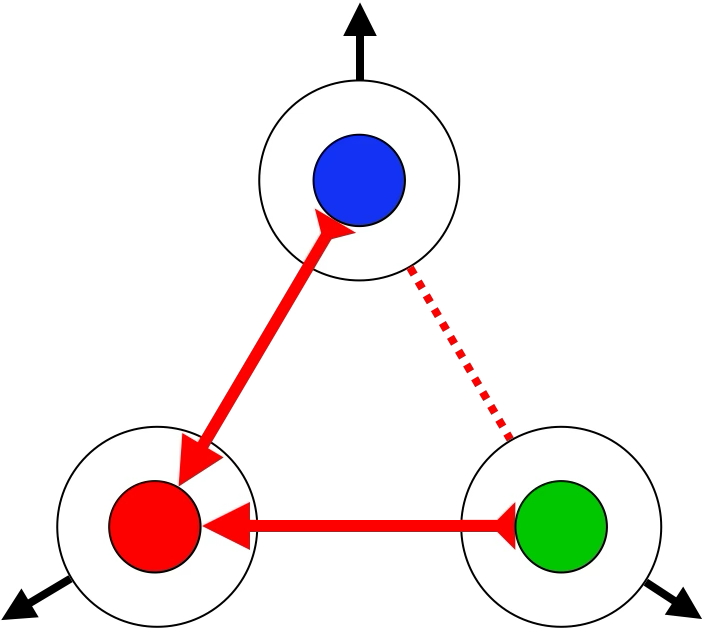}
\caption{Triangle network and partition recovery.
\emph{Left to right:}~Healthy triangle (orange stars mark active
links); $b$--$c$ partitions, recovery through~$a$ (blue);
$a$--$c$ partitions, recovery through~$b$ (green);
$a$--$b$ partitions, recovery through~$c$ (red).}
\label{fig:triangle-recovery}
\end{figure}

Gray's two-phase commit protocol~\citep{gray1978-2pc} suffers from a
well-known blocking vulnerability: if the coordinator fails after
sending \textsc{prepare} but before sending \textsc{commit}, all
participants are stuck in the \textsc{tentative} state with no way
to resolve.  Three-phase commit~\citep{skeen1981} addresses this by
adding a pre-commit phase, but at the cost of additional message
rounds and continued vulnerability to certain failure patterns.%

The triangle network eliminates blocking by eliminating the
coordinator.  In a triangle of three nodes $\{A, B, C\}$, every pair
is connected by a bilateral \oae{} link.  If the $A$--$B$ link
partitions during a transaction, nodes $A$ and $C$ can still
communicate, as can $B$ and $C$.  The third node $C$ serves as a
\emph{recovery resource}: it can relay the reflecting phase through
the $A$--$C$--$B$ path, allowing the transaction to complete without
a coordinator and without waiting for the partitioned link to heal.%

\begin{definition}[Triangle Consistency Property]
\label{def:triangle}
In a network where every pair of nodes is connected through at least
one triangle, any bilateral transaction can complete its reflecting
phase through an alternative path whenever a direct link is
partitioned.  The network maintains semantic consistency as long as
at least one path between any pair of nodes remains operational.
\end{definition}

\subsection{Octavalent Substrate}

\begin{figure}[h]
\centering
\includegraphics[width=0.8\columnwidth]{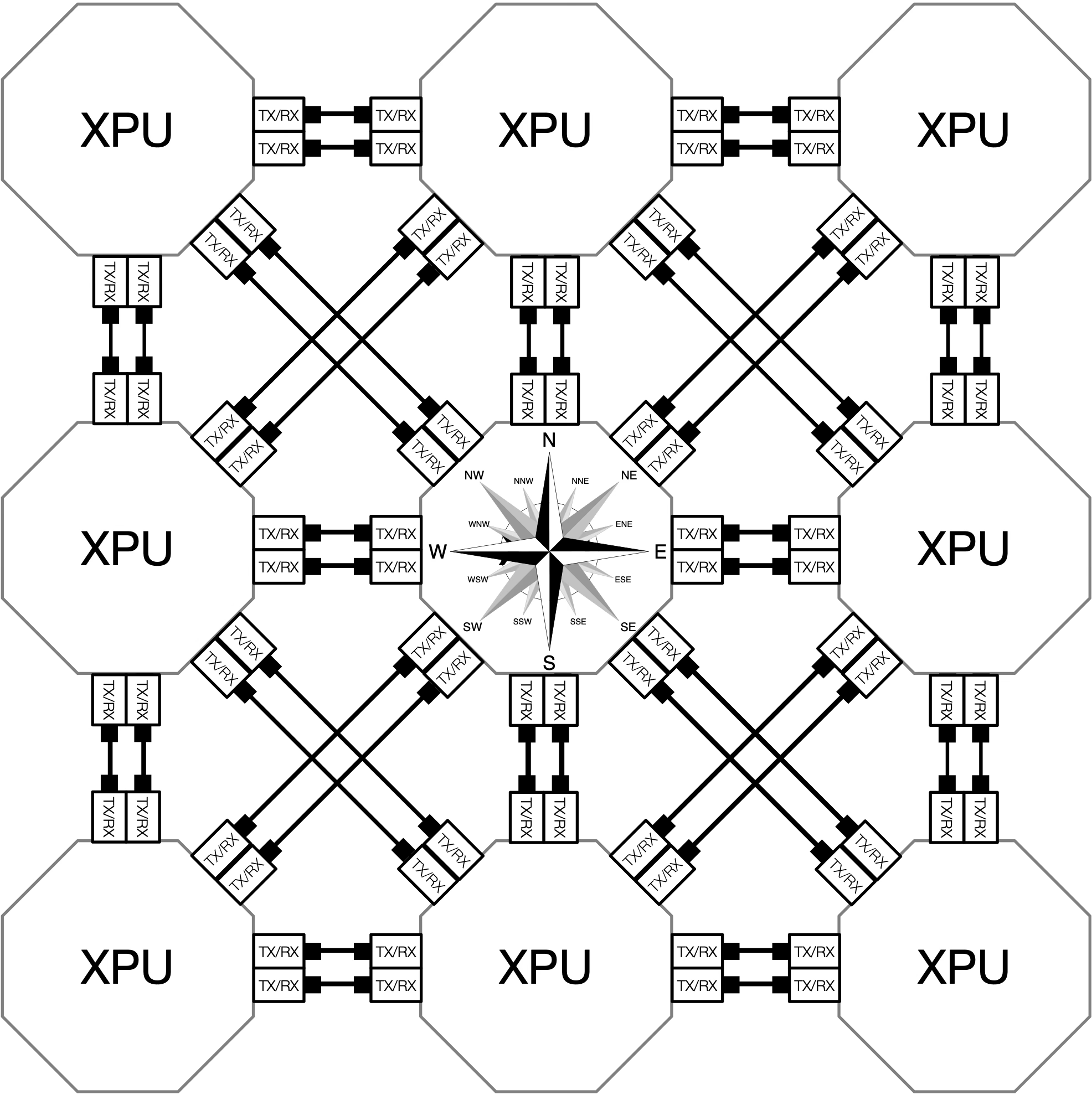}
\caption{The octavalent topology.  Each cell contains an XPU
with eight bilateral TX/RX ports (compass directions N through NW).
Every adjacent pair participates in multiple triangles, providing
massive redundancy for the reflecting path.}
\label{fig:octavalent}
\end{figure}

The \oae{} physical architecture uses an octavalent topology: each
cell (module server plus 8-port NIC) connects to eight neighbors in
a 2D plane via PCB traces or short cables ($<$0.5\,m).  In this
topology, every pair of adjacent nodes participates in multiple
triangles, providing massive redundancy for the reflecting path.%

Nodes and links are dynamically reconfigurable---not only on failure
and healing events, but as microservice traffic patterns elastically
expand and shrink.  Each cell builds a directed acyclic graph (DAG)
on each port, providing the acyclic property required for healing
with local information only.

\subsection{Scale Independence}

The triangle property is scale-independent: it applies identically
to three nodes on a chiplet, three racks in a data center, and three
data centers across a continent.  The only variable is the temporal
horizon of the reflecting phase---the time within which a reflection
must arrive for the transaction to commit.  At chiplet scale, this
horizon is nanoseconds.  At continental scale, it is milliseconds.
The protocol structure is identical; only the clock rate changes.

This is the engineering realization of the \rcp{}: the reversible
causal principle is scale-invariant because it depends on the
\emph{structure} of the exchange (bilateral, reflecting,
information-conserving), not on the \emph{speed} of the medium.

\section[Nine Sublayers]{The Nine-Sublayer \oae{} Architecture (L2.1--L2.9)}
\label{sec:nine-layers}

The full \oae{} architecture~\citep{borrill2025-oae9} organizes the
Leibniz Bridge into nine sublayers, each implementing a specific
aspect of mutual information conservation.  A critical point of
orientation: \textbf{all nine sublayers reside entirely within what
the ISO/OSI model calls Layer~2} (Data Link).  The \oae{} stack is
not a replacement for the OSI model; it is a fine-grained
decomposition of the bilateral transaction machinery that fits inside
a single Ethernet frame, below where IP begins.  We label the
sublayers L2.1--L2.9 to make this nesting explicit:

\begin{center}
\small
\begin{tabular}{cll}
\toprule
\textbf{Sublayer} & \textbf{Name} & \textbf{Conservation role} \\
\midrule
L2.9 & Application & Semantic invariants \\
L2.8 & Transaction & Atomicity boundaries \\
L2.7 & Session & Cross-transaction state \\
L2.6 & Agreement & Bilateral commitment \\
L2.5 & Reflection & Knowledge balance verification \\
L2.4 & Transport & Reliable delivery \\
L2.3 & Network & Routing and topology \\
L2.2 & Link & Bilateral frame exchange \\
L2.1 & Physical & Bidirectional signaling \\
\bottomrule
\end{tabular}
\end{center}

The names deliberately echo the OSI layers (Physical, Link, Network,
Transport, Session, Application) to emphasize the structural
parallel---but the numbering L2.$n$ is a constant reminder that the
entire stack operates within the Ethernet frame.  What the OSI model
handles across seven layers and multiple protocol boundaries, the
\oae{} architecture handles within a single Layer~2 PDU by making the
bilateral transaction structure explicit at every sublayer.

The critical architectural difference is the presence of L2.5
(Reflection) and L2.6 (Agreement) as \emph{distinct, mandatory
sublayers}.  No conventional Layer~2 protocol includes these.  In
the OSI model, reflection is optional and agreement is delegated to
the application (Layer~7).  In the \oae{} model, they are
architectural requirements at the \emph{link} level: no transaction
can cross a sublayer boundary without completing its reflecting phase
and verifying knowledge balance.%

The sublayered structure composes: each sublayer can be treated as a
\pif{} link with its own forward and reverse operators, and the
composition of sublayers preserves the information conservation
invariant.  This gives the architecture a \emph{dagger-monoidal}
categorical structure: series composition ($\odot$) captures
sublayer stacking, parallel composition ($\otimes$) captures
spatial distribution, and the dagger operation captures time
reversal~\citep{munamala2025}.

\section[Open Questions]{Open Questions}
\label{sec:open}

The Leibniz Bridge opens more questions than it closes.  We identify
seven that we consider most pressing.

\paragraph{1.\ Can the semantic arrow be formalized in process algebra?}
The categorical framework (dagger-monoidal categories, process
matrices) provides the mathematical vocabulary, but a complete
process algebra for semantic transactions---with composition,
hiding, and refinement operators---has not yet been constructed.
The Abramsky--Gorard categorical semantics framework offers a
promising starting point.

\paragraph{2.\ What is the relationship to Hardy's causal structure?}
Hardy's framework for quantum gravity~\citep{hardy2001} replaces
fixed background spacetime with a dynamic causal structure.  The
\oae{}'s dynamic DAG topology---where causal paths are created and
destroyed by the protocol rather than assumed by the model---may be
a classical implementation of Hardy's vision.  The connection
deserves formal investigation.

\paragraph{3.\ Does quantum indefinite causal order provide a
physical substrate?}
Zhao et al.~\citep{zhao2025} showed that \ico{} provides a genuine
communication advantage.  If bilateral Ethernet links implement a
classical form of \ico{}, does this advantage extend to classical
networking?  The \pif{} capacity result
($C_{\pif{}} = 2 \cdot C_{\text{one-way}}$) suggests it does, but a
rigorous proof linking the quantum and classical cases is needed.

\paragraph{4.\ Can LLMs be trained with semantic transaction
guarantees?}
Part~IV showed that autoregressive generation is \fito{} at the
token level.  Could a training architecture that incorporates a
reflecting phase---where each generated token is verified against
semantic constraints before being committed to the output---reduce
hallucination?  Self-consistency methods~\citep{wang2022} approximate
this at the sequence level, but token-level reflection remains
unexplored.

\paragraph{5.\ Edward Lee's Lingua Franca and the Maxwait mechanism.}
Lee's Lingua Franca framework~\citep{lohstroh2021} forces programmers
to state timing assumptions explicitly and mandates fault handlers
for when those assumptions are violated.  The Maxwait
mechanism---waiting for the slowest participant---may already
implement a form of bilateral causal exchange.  If Maxwait preserves
phase-locking between participants, it may implicitly enforce a
condition similar to \pif{}'s $R \simeq F^{-1}$.  This needs careful
analysis.

\paragraph{6.\ Ken Birman's Virtual Synchrony in the \ico{} framework.}
Birman's Virtual Synchrony~\citep{birman1987} is provably optimal
\emph{within the definite causal order constraint}.  What happens when
the constraint is relaxed?  Tensor Clocks from the Indefinite Logical
Timestamps framework can represent the full 2-bit state space.  Does
this change the optimality landscape for group communication
protocols?

\paragraph{7.\ Reformulating the Alternating Causality $dI/dt$
framework.}
Lee~\citep{lee-email2026} identified a potential category error in the
Alternating Causality formulation: Shannon's mutual information
$I(X;Y)$ is a property of a joint probability distribution, not a
function of time, so the derivatives $dI_+/dt$ and $dI_-/dt$ are
individually zero---not just in sum.  Three possible reformulations
exist: (a)~interpret the notation as channel capacity \emph{utilization
rates} (bits/second allocated to each direction), not as derivatives of
mutual information; (b)~move to time-indexed stochastic processes where
the joint distribution $P(X_t, Y_t)$ itself evolves, making $I_t$ a
legitimate function of time; or (c)~abandon the dynamical notation
entirely and express \pif{} as a static constraint on bilateral
capacity allocation.  Resolving this is critical: if the formalism
conflates information-theoretic statics with dynamical equations, the
AC framework requires reformulation on sound mathematical footing.

\section[The Mulligan Stew]{Coda: The Mulligan Stew}
\label{sec:coda}

The Mulligan Stew is a weekly meeting of physicists, computer
scientists, and engineers---the ``proving ground'' for the ideas in
this series.  For over two years, this group has wrestled with the
questions that these five papers attempt to answer.  The name
reflects the diverse mix of expertise: everyone brings what they
have, and the result is richer than any individual ingredient.

Pat Helland, whose foundational contributions to distributed
transaction processing span four decades---from Tandem's NonStop to
Microsoft's distributed SQL---once expressed a wish for a ``do
over'': the opportunity to rebuild distributed computing with the
benefit of everything we have learned since the 1970s about what
goes wrong and why.%

This series is that do-over.

Not because it has all the answers---the open questions in
Section~\ref{sec:open} are genuine, and several may take years to
resolve.  But because it identifies the single structural error that
produced the need for a do-over in the first place: the assumption
that time flows forward, that forward flow implies semantic progress,
and that the return path is overhead rather than the constitutive
mechanism through which meaning is established.

The Leibniz Bridge replaces this assumption with a principle: mutual
information conservation.  The semantic arrow of time is not assumed;
it is \emph{constructed}, exchange by exchange, reflection by
reflection, commitment by commitment.  The direction of time on a
link is not a physical law---it is an emergent property of the
entropy produced when a reversible exchange irreversibly commits.

Edward Ashford Lee put it most precisely: ``There's a split between
reality and models here, and you got to keep that split
clear.''~\citep{lee2026}  The \fito{} category mistake collapses
that split.  The Leibniz Bridge restores it.

The Mulligan Stew continues.



\begin{thebibliography}{99}
\bibitem[Lee(2026, private)]{lee-email2026}
\citenamefont{Lee}, Edward~A.
Private communication to Paul Borrill, 02026-FEB-09.
On the $dI/dt$ formulation: ``In Shannon's formulation, mutual
information is not a function of time, so these derivatives are
individually zero, so not just the sum is zero.''
\end{thebibliography}
\end{document}